*International*

*Virtual*

*Observatory*

*Alliance*

# IVOA Astronomical Data Query Language

# Version 2.0

## *IVOA Recommendation 30 Oct 2008*

**Editor(s):**
Pedro Osuna and Inaki Ortiz
**Author(s):**
Inaki Ortiz, Jeff Lusted, Pat Dowler, Alexander Szalay, Yuji Shirasaki, Maria A. Nieto-Santisteban, Masatoshi Ohishi, William O'Mullane, Pedro Osuna, the VOQL-TEG and the VOQL Working Group.

---

## Abstract

This document describes the Astronomical Data Query Language (ADQL). ADQL has been developed based on SQL92. This document describes the subset of the SQL grammar supported by ADQL. Special restrictions and extensions to SQL92 have been defined in order to support generic and astronomy specific operations.

## Status of This Document

This document has been produced by the VO Query Language Working Group.

It has been reviewed by IVOA Members and other interested parties, and has been endorsed by the IVOA Executive Committee as an IVOA Recommendation. It is a stable document and may be used as reference material or cited as a normative reference from another document. IVOA's role in making the Recommendation is to draw attention to the specification and to promote its widespread deployment. This enhances the functionality and interoperability inside the Astronomical Community.

Please note that version 1.0 was never promoted to Proposed Recommendation.

*A list of current IVOA Recommendations and other technical documents can be found at http://www.ivoa.net/Documents/.*

## Contents

# 1 Introduction

The Astronomical Data Query Language (ADQL) is the language used by the International Virtual Observatory Alliance (IVOA) to represent astronomy queries posted to VO services. The IVOA has developed several standardized protocols to access astronomical data, e.g., SIAP and SSAP for image and spectral data respectively. These protocols might be satisfied using a single table query. However, different VO services have different needs in terms of query complexity and ADQL arises in this context.

The ADQL specification makes no distinction between core and advanced or extended functionalities. Hence ADQL has been built according to a single language definition (BNF based [1]). Any service making use of ADQL would then define the level of compliancy to the language. This would allow the notion of core and extension to be service-driven and it would decouple the language from the service specifications.

ADQL is based on the Structured Query Language (SQL), especially on SQL 92. The VO has a number of tabular data sets and many of them are stored in relational databases, making SQL a convenient access means. A subset of the SQL grammar has been extended to support queries that are specific to astronomy. Similarly to SQL, the ADQL language definition is not semantically safe by design and therefore this specification defines syntactical correctness only. Type safety has been achieved as far as it can be done in SQL.

The exact meaning of key words indicating requirement levels can be found in the References section [2].

# 2 Astronomical Data Query Language (ADQL)

This section describes the ADQL language specification. We will define in subsequent sections the syntax for the special characters, reserved and non-reserved words, identifiers and literals and then, finally, the syntax for the query expression.

The formal notation for syntax of computing languages is often expressed in the “Backus Naur Form” BNF. This syntax is used by popular tools for producing parsers. Appendix A to this document provides the full BNF grammar for ADQL.

The following conventions are used through this document:

- Optional items are enclosed in meta symbols [ and ]
- A group of items is enclosed in meta symbols { and }
- Repetitive item (zero or more times) are followed by …
- Terminal symbols are enclosed by < and >

- Terminals of meta-symbol characters (=,[,],(,),<,>,*) are surrounded by quotes (“) to distinguish them from meta-symbols
- Case insensitiveness otherwise stated.

## 2.1 Characters, Keywords, Identifiers and Literals

### 2.1.1 Characters

The language allows simple Latin letters (lower and upper case, i.e. {aA-zZ}), digits ({0-9}) and the following special characters:

- space
- single quote (’)
- double quote (“)
- percent (%)
- left and right parenthesis
- asterisk (*)
- plus sign (+)
- minus sign (-)
- comma (,)
- period (.)
- solidus (/)
- colon (:)
- semicolon (;)
- less than operator (<)
- equals operator (=)
- greater than operator (>)
- underscore (_)
- ampersand (&)
- question mark (?)
- vertical bar (|)

### 2.1.2 Keywords and Identifiers

Besides the character set, the language provides a list of reserved keywords plus the syntax description for regular identifiers.

A reserved keyword has a special meaning in ADQL and cannot be used as an identifier. These keywords must be enforced and should be extensive as an escaping mechanism is already in place. We can extend the list of SQL92 reserved keywords to accommodate those useful for astronomical purposes and/or present in a subset of vendor specific languages only (e.g. TOP). This leads to the following list:

- SQL reserved keywords:

ABSOLUTE, ACTION, ADD, ALL, ALLOCATE, ALTER, AND, ANY, ARE, AS, ASC, ASSERTION, AT, AUTHORIZATION, AVG, BEGIN, BETWEEN, BIT, BIT_LENGTH, BOTH, BY, CASCADE, CASCADED, CASE, CAST, CATALOG, CHAR, CHARACTER, CHARACTER_LENGTH, CHAR_LENGTH, CHECK, CLOSE, COALESCE, COLLATE, COLLATION, COLUMN, COMMIT, CONNECT, CONNECTION, CONSTRAINT, CONSTRAINTS, CONTINUE, CONVERT, CORRESPONDING, COUNT, CREATE, CROSS, CURRENT, CURRENT_DATE, CURRENT_TIME, CURRENT_TIMESTAMP, CURRENT_USER, CURSOR, DATE, DAY, DEALLOCATE, DECIMAL, DECLARE, DEFAULT, DEFERRABLE, DEFERRED, DELETE, DESC, DESCRIBE, DESCRIPTOR, DIAGNOSTICS, DISCONNECT, DISTINCT, DOMAIN, DOUBLE, DROP, ELSE, END, END-EXEC, ESCAPE, EXCEPT, EXCEPTION, EXEC, EXECUTE, EXISTS, EXTERNAL, EXTRACT, FALSE, FETCH, FIRST, FLOAT, FOR, FOREIGN, FOUND, FROM, FULL, GET, GLOBAL, GO, GOTO, GRANT, GROUP, HAVING, HOUR, IDENTITY, IMMEDIATE, IN, INDICATOR, INITIALLY, INNER, INPUT, INSENSITIVE, INSERT, INT, INTEGER, INTERSECT, INTERVAL, INTO, IS, ISOLATION, JOIN, KEY, LANGUAGE, LAST, LEADING, LEFT, LEVEL, LIKE, LOCAL, LOWER, MATCH, MAX, MIN, MINUTE, MODULE, MONTH, NAMES, NATIONAL, NATURAL, NCHAR, NEXT, NO, NOT, NULL, NULLIF, NUMERIC, OCTET_LENGTH, OF, ON, ONLY, OPEN, OPTION, OR, ORDER, OUTER, OUTPUT, OVERLAPS, PAD, PARTIAL, POSITION, PRECISION, PREPARE, PRESERVE, PRIMARY, PRIOR, PRIVILEGES, PROCEDURE, PUBLIC, READ, REAL, REFERENCES, RELATIVE, RESTRICT, REVOKE, RIGHT, ROLLBACK, ROWS, SCHEMA, SCROLL, SECOND, SECTION, SELECT, SESSION, SESSION_USER, SET, SIZE, SMALLINT, SOME, SPACE, SQL, SQLCODE, SQLERROR, SQLSTATE, SUBSTRING, SUM, SYSTEM_USER, TABLE, TEMPORARY, THEN, TIME, TIMESTAMP, TIMEZONE_HOUR, TIMEZONE_MINUTE, TO, TRAILING, TRANSACTION, TRANSLATE, TRANSLATION, TRIM, TRUE, UNION, UNIQUE, UNKNOWN, UPDATE, UPPER, USAGE, USER, USING, VALUE, VALUES, VARCHAR, VARYING, VIEW, WHEN, WHENEVER, WHERE, WITH, WORK, WRITE, YEAR, ZONE

- ADQL reserved keywords:

ABS, ACOS, ASIN, ATAN, ATAN2, CEILING, COS, DEGREES, EXP, FLOOR, LOG, LOG10, MOD, PI, POWER, RADIANS, RAND, ROUND, SIN, SQRT, TAN, TOP, TRUNCATE

The identifiers are used to express, for example, a table or a column reference name.

Both the identifiers and the keywords are case insensitive. They SHALL begin with a letter {aA-zZ}. Subsequent characters shall be letters, underscores or digits {0-9} as follows:

<Latin_letter> [{ <underscore> | {<Latin_letter> | <digit>} }]

For practical purposes the language specification should be able to address reserved keyword and special character conflicts. To do so the language provides a way to escape a non-compliant identifier by using the double quote character as a delimiter.

ADQL allows making use of the same quoting mechanism to handle the case sensitiveness if needed.

### 2.1.3 Literals

Finally we define the syntax rules for the different data types: string and number.

A string literal is a character expression delimited by single quotes.

Literal numbers are expressed in BNF as follows:

```
<unsigned_numeric_literal> ::=
        <exact_numeric_literal> | <approximate_numeric_literal>
<exact_numeric_literal> ::=
        <unsigned_integer> [<period> [<unsigned_integer>]]
        | <period><unsigned_integer>
<unsigned_integer> ::= <digit>…
<approximate_numeric_literal> ::= <mantissa> E <exponent>
<mantissa> ::= <exact_numeric_literal>
<exponent> ::= <signed_integer>
<signed_integer> ::= [<sign>] <unsigned_integer>
<sign> ::= <plus_sign> | <minus_sign>
```

Regarding the usage of other data types like datetime and timestamp, ADQL can deal with them similarly to how SQL does: using the string literal construct. As Relation Database Manager Systems (RDBMs) do, a service should be able to implicitly convert strings to internal (datetime or timestamp) form using a variety of techniques, where e.g. ISO 8601 is an acceptable format. Therefore, as with other string representations, it should be up to the service capability to understand such specific formats.

## *2.2 Query syntax*

A full and complete syntax of the select statement can be found in “Appendix A: BNF Grammar” at the <query_specification> construct. Follows a simplified syntax for the SELECT statement showing the main constructs for the query specification:

```
SELECT [ ALL |  DISTINCT ]
        [ TOP unsigned_integer ]
        { * | { value_expression [ [AS] column_name ] }, …  }
        FROM {
          { table_name [ [AS] identifier ]      |
            ( SELECT ….) [ [AS] identifier ]  |
            table_name [NATURAL] [ INNER | { LEFT | RIGHT | FULL
              [OUTER] } ] JOIN table_name
              [ON search_condition | USING ( column_name,…) ] }
        , …}
        [ WHERE search_condition ]
        [ GROUP BY column_name, … ]
        [ HAVING search_condition  ]
        [ ORDER BY { column_name | unsigned_integer } [ ASC | DESC
            ],…]
```

The SELECT statement defines a query to some derived table(s) specified in the FROM clause. As a result of this query, a subset of the table(s) is returned. The order of the rows MAY be arbitrary unless ORDER BY clause is specified. The order of the columns to return SHALL be the same as the order specified in the selection list, or the order defined in the original table if asterisk is specified.

TOP n construct is used to return the first n-rows.

The selection list MAY include any numeric, string or geometry value expression.

In the following sections some constructs requiring further description are presented.

### 2.2.1 Table subqueries and Joins

Table subqueries are present and can be used by some existing predicates within the search condition (IN and BETWEEN most likely) or as an artifact of building derived tables.

Among the different types of join, ADQL supports INNER and OUTER (LEFT, RIGHT and FULL) joins. If none is specified, the default is INNER. All of these can be NATURAL or not. The join condition does not support embedded sub joins.

### 2.2.2 Search condition

The search condition can be part of several other clauses: JOIN, HAVING and, obviously, WHERE. Standard logical operators are present in its description (AND, OR and NOT). Five different types of predicates are present in which different types of reserved keywords or characters are used:

- Standard comparison operators: =, !=, <>, <, >, <=, >=
- BETWEEN
- LIKE
- NULL
- EXISTS

## 2.3 Mathematical and Trigonometrical Functions

ADQL declares a list of reserved keywords (section 2.1.2) which defines a set of mathematical and trigonometrical function names. Their syntax, usage and description are detailed in the following tables:

| Name | Argument data type | Return data type | Description |
|---|---|---|---|
| abs(x) | double | double | Returns the absolute value of *x*. |
| ceiling(x) | double | double | Returns the smallest double value that is not less than the argument *x* and is equal to a mathematical integer. |
| degrees(x) | double | double | Converts an angle to degrees. Argument *x* must be in radians. |
| exp(x) | double | double | Returns Euler's number e raised to the power of *x*. |
| floor(x) | double | double | Returns the largest double value that is not greater than the argument *x* and is equal to a mathematical integer. |
| log(x) | double | double | Returns the natural logarithm (base e) of a double value. Value *x* must be greater than zero. |
| log10(x) | double | double | Returns the base 10 logarithm of a double value. Value *x* must be greater than zero. |
| mod(x, y) | double | double | Returns the remainder of *y*/*x*. |
| pi() | n/a | double | The π constant. |
| power(x, y) | *x* double<br>*y* double | double | Returns the value of the first argument raised to the power of the second argument. |
| radians(x) | double | double | Converts an angle to radians. Argument *x* must be in degrees. |
| sqrt(x) | double | double | Returns the positive square root of a double value. |
| rand(x) | integer | double | Returns a random value between 0.0 and 1.0, where *x* is a seed value. |
| round(x, n) | *x* double<br>*n* integer | double | Rounds double value *x* to *n* number of decimal places, with the default being to round to the nearest integer. To round to the left of the decimal point, a negative number should be provided. |
| truncate(x, n) | *x* double<br>*n* integer | double | Returns the result of truncating the argument *x* to *n* decimal places. |

**Table 1: Mathematical functions**

| Name | Argument data type | Return data type | Description |
|---|---|---|---|
| acos(x) | double | double | Returns the arc cosine of an angle, in the range of 0 through π radians. Absolute value of *x* must be lower or equal than 1.0. |
| asin(x) | double | double | Returns the arc sine of an angle, in the range of -π/2 through π/2 radians. Absolute value of *x* must be and lower or equal than 1.0. |
| atan(x) | double | double | Returns the arc tangent of an angle, in the range of -π/2 through π/2 radians. |
| atan2(y,x) | double | double | Converts rectangular coordinates x,y to polar angle. It computes the arc tangent of *y/x* in the range of –π through π radians. |
| cos(x) | double | double | Returns the cosine of an angle, in the range of -1.0 through 1.0. Argument *x* must be in radians. |
| sin(x) | double | double | Returns the sine of an angle, in the range of -1.0 through 1.0. Argument *x* must be in radians. |
| tan(x) | double | double | Returns the tangent of an angle. Argument *x* must be in radians. |

**Table 2: Trigonometrical functions**

## *2.4 Geometrical Functions*

### 2.4.1 Overview

In addition to the mathematical functions, ADQL provides a set of geometrical functions to enhance the astronomical usage of the language. The list of ADQL reserved keywords shown in Section 2.1.2 is therefore extended with the following function names:

AREA, BOX, CENTROID, CIRCLE, CONTAINS, COORD1, COORD2, COORDSYS, DISTANCE, INTERSECTS, POINT, POLYGON, REGION

A special attention has to be paid to the REGION function. As can be seen more in detail in Section 2.4.14, this construct is a general purpose function and it takes a string value expression as argument. The format of the string is to be specified by a service that accepts ADQL by referring to a standard format. Currently STC/s (See [3] and [4]) is the only standardized string representation a service can declare.

As can also be seen in the following sections, all these functions have arguments being a geometrical, a string and/or a numerical value expression. When these values represent spherical coordinates the units MUST be in degrees (square degrees for area). If the cartesian coordinate system is used, the vector coordinates MUST be normalized.

Regarding the legal ranges, for spherical coordinates, these SHOULD be [0, 360] and [-90, 90]. In a cartesian coordinate system, there are no inherent limits but the already mention constrain that vectors should be normalized. It remains up to the service making use of ADQL to define the errors that should be raised when using values outside these ranges.

For all these functions there is also a string parameter defining the coordinate system. The allowed values MUST be defined by any service making use of ADQL. A list of standard coordinate system literals can be found in the appendix of the STC specification [3].

Generally speaking, all these geometrical functions cover three different topics: data types, predicates and utility calculations. Each of these are covered below.

### Data Type Functions

Certain functions represent geometry data types. These data types are BOX, CENTROID, CIRCLE, POINT and POLYGON together with the generalized REGION data type. The functions are similarly named and return a variable-length binary value. The semantics of these data types are based on the corresponding concepts from the STC data model (See [3]).

Geometry data types are centered around the BNF construct value_expression which is central to data types within SQL.

```
<value_expression> ::=
     <numeric_value_expression>
   | <string_value_expression>
   | <geometry_value_expression>
```

A <geometry_value_expression> does not simply cover data type functions (POINT, CIRCLE etc) but must also allow for column values where a geometry data type is stored in a column. Therefore, <geometry_value_expression> is expanded as

```
<geometry_value_expression> ::= <value_expression_primary> |
   <geometry_value_function>
```

, where

```
<geometry_value_function> ::= <box> | <centroid> | <circle> | <point> |
    <polygon> | <region>
```

and

<value_expression_primary> makes possible to use a column reference.

### Predicate Functions

Functions CONTAINS and INTERSECTS each accept two geometry data types and return 1 or 0 according to whether the relevant verb (e.g.: "contains") is satisfied against the two input geometries; 1 represents true and 0 represents false. Each of these functions can be assembled into a predicate:

```
SELECT * FROM SDSS as s WHERE CONTAINS(POINT(...), CIRCLE(...)) = 1
```

, where the ... would represent the constituent parts of a CIRCLE and POINT geometry.

One would expect later additions to ADQL to add to this range of functions. For example: equals, disjoint, touches, crosses, within, overlaps and relate are possibilities.

### Utility Functions

Function COORDSYS extracts the coordinate system string from a given geometry. To do so it accepts a geometry expression and returns a calculated string value.

This function has been included as a string value function because it returns a simple string value. Hence

```
<string_value_function> :: = <string_geometry_function>
  | <user_defined_function>

<string_geometry_function> ::=  <extract_coordsys>

<extract_coordsys> ::= COORDSYS <left_paren>
  <geometry_value_expression> <right_paren>
```

Other functions like AREA, COORD1, COORD2 and DISTANCE accept a geometry (or two geometries in the case of DISTANCE) and return a calculated numeric value.

The Predicate and most of the Utility functions have been included as numeric value functions because they return simple numeric values. Thus

```
<numeric_value_function> ::=
   <trig_function>
 | <math_function>
 | <numeric_geometry_function>
 | <user_defined_function>
```

, where

<numeric_geometry_function> ::=  <predicate_geometry_function>
  | <non_predicate_geometry_function>

and

<non_predicate_geometry_function> ::=
  AREA <left_paren> <geometry_value_expression> <right_paren>
  | COORD1 <left_paren> <coord_value> <right_paren>
  | COORD2 <left_paren> <coord_value> <right_paren>
  | DISTANCE <left_paren> <coord_value> <comma> <coord_value>
    <right_paren>

and

<predicate_geometry_function> ::= <contains> | <intersects>

The following sections provide a detailed description for each geometrical function. Its functionality and usage is described rather than going into the BNF grammar details as above.

### 2.4.2 AREA

This function computes the area, in square degrees, of a given geometry.

For example, the area of a circle of one degree radius centered in a position of (25.4, -20.0) degrees and defined according to the ICRS coordinate system with GEOCENTER reference position would be written as follows:

AREA(CIRCLE('ICRS GEOCENTER', 25.4, -20.0, 1))

The coordinates of the circle center could also be directly derived from either a POINT function (See 2.4.12) or the coordinate's column references:

AREA(CIRCLE('ICRS GEOCENTER', t.ra, t.dec, 1))

, where *t* would be the table and *ra*, *dec* the column references for the circle centre.

Inappropriate geometries for this construct (e.g. POINT) SHOULD either return zero or throw an error message, the later to be defined by the service making use of ADQL.

### 2.4.3 BOX

This function expresses a box on the sky. A box is a special case of Polygon, defined purely for convenience, and it corresponds semantically to the STC Box region ([3], section 4.5.1.5). It is specified by a center position and size (in both coordinates) defining a cross centered on the center position and with arms extending, parallel to the coordinate axes at the center position, for half the respective sizes on either side. The box's sides are line segments or great circles intersecting the arms of the cross in its end points at right angles with the arms.

The function arguments specify the coordinate system, the center position and both the width and height (arms) values, where

- the coordinate system is a string value expression as defined in Section 2.4.1.
- the center position is a comma separated numeric duple, with units and legal ranges as defined in Section 2.4.1.
- and the arms are numeric value expressions in degrees.

For example, a function expressing a box of ten degrees centered in a position (25.4, -20.0) in degrees and defined according to the ICRS coordinate system with GEOCENTER reference position would be written as follows:

```
BOX('ICRS GEOCENTER', 25.4, -20.0, 10, 10)
```

As another example, the coordinates of the center position could also be extracted from either a POINT function (See 2.4.12) or the coordinate's column references:

```
BOX('ICRS GEOCENTER', t.ra, t.dec, 10, 10)
```

, where *t* would be the table and *ra*, *dec* the column references for the center position.

To see what this function would return when listed in the select clause, see Section 2.4.15.

### 2.4.4 CENTROID

This function computes the centroid of a given geometry and returns a POINT (See 2.4.11).

For example, the centroid of a circle of one degree radius centered in a position of (25.4, -20.0) degrees and defined according to the ICRS coordinate system with GEOCENTER reference position would be written as follows :

```
CENTROID(CIRCLE ('ICRS GEOCENTER', 25.4, -20.0, 1))
```

### 2.4.5 CIRCLE

This function expresses a circular region on the sky (a cone in space) and corresponds semantically to the STC Circle region ([3], section 4.5.1.2).. The function arguments specify the coordinate system, the center position, and the radius, where:

- the coordinate system is a string value expression as defined in Section 2.4.1.
- the center position is a comma separated numeric duple, with units and legal ranges as defined in Section 2.4.1.
- and the radius is a numeric value expression in degrees.

For example, a function expressing a circle of one degree radius centered in a position of (25.4, -20.0) degrees and defined according to the ICRS coordinate system with GEOCENTER reference position, would be written as follows:

```
CIRCLE('ICRS GEOCENTER', 25.4, -20.0, 1)
```

The coordinates of the center position could also be derived from either a POINT function (See 2.4.12) or the coordinate's column references:

```
CIRCLE('ICRS GEOCENTER', t.ra, t.dec, 1)
```

, where *t* would be the table and *ra*, *dec* the column references for the center position.

To see what this function would return when listed in the select clause, see Section 2.4.15.

### 2.4.6 CONTAINS

This numeric function determines if a geometry is wholly contained within another. This is most commonly used to express the "point-in-shape" condition.

For example, to determine if a point with right ascension of 25 degrees and declination of -19.5 degrees according to the ICRS coordinate system with GEOCENTER reference position is within a circle of one degree radius centered in a position of (25.4, -20.0) degrees and defined according to the same coordinate system, we would make use of the CONTAINS function as follows:

```
CONTAINS(POINT('ICRS GEOCENTER', 25.0,-19.5),
         CIRCLE('ICRS GEOCENTER', 25.4, -20.0, 1))
```

, where the CONTAINS function returns 1 (true) if the first argument is in or on the boundary of the circle and 0 (false) otherwise. Thus, contains is not symmetric in the meaning of the arguments. When used in the where clause of a query, the value must be compared to 0 or 1 to form an SQL predicate:

CONTAINS(POINT('ICRS GEOCENTER', 25.0,-19.5),
CIRCLE('ICRS GEOCENTER', 25.4, -20.0, 1)) = 1

for "does contain" and

CONTAINS(POINT('ICRS GEOCENTER', 25.0,-19.5),
CIRCLE('ICRS GEOCENTER', 25.4, -20.0, 1)) = 0

for "does not contain".

The arguments to the CONTAINS function can be (literal) values created from the geometry types or they can be single column names or aliases (for geometry stored in a database table). Since the two argument geometries may be expressed in different coordinate systems, the function is responsible for converting one (or both). If it cannot do so, it SHOULD throw an error message, to be defined by the service making use of ADQL.

### 2.4.7 COORD1

This function extracts the first coordinate value, in degrees, of a given POINT (See 2.4.12) or column reference.

For example, the right ascension of a point with position (25, -19.5) in degrees according to the ICRS coordinate system with GEOCENTER reference position, would be obtained using the following expression:

COORD1(POINT('ICRS GEOCENTER', 25.0,-19.5))

, being the result a numeric value of 25.0 degrees. The fist coordinate could also be derived directly from a column reference as follows:

COORD1(t.point)

, where *t* is the table and *point* the column reference for the POINT geometry stored in the database table.

### 2.4.8 COORD2

This function extracts the second coordinate value, in degrees, of a given POINT (See 2.4.12) or column reference.

For example, the declination of a point with position (25, -19.5) in degrees according to the ICRS coordinate system with GEOCENTER reference position, would be obtained using the following expression:

COORD2(POINT('ICRS GEOCENTER', 25.0,-19.5))

, being the result a numeric value of -19.5 degrees. The second coordinate could also be derived directly from a column reference as follows:

COORD2(t.point)

, where *t* is the table and *point* the column reference for the POINT geometry stored in the database table.

### 2.4.9 COORDSYS

This function extracts the coordinate system string value from a given geometry. As described in section 2.4.1, the allowed return values must be defined by any service making use of ADQL, and a list of standard coordinate system literals can be found in the STC specification [3].

For example, a function extracting the coordinate system of a point with position (25, -19.5) in degrees according to the ICRS coordinate system with GEOCENTER reference position, would be written as follows:

COORDSYS(POINT('ICRS GEOCENTER', 25.0,-19.5))

, returning the 'ICRS GEOCENTER' string literal. As other samples above, the coordinate system could also be derived from a column referencing any other geometry data type:

COORDSYS(t.circle)

, where *t* is the table and *circle* the column reference for the CIRCLE geometry stored in the database table.

### 2.4.10 DISTANCE

This function computes the arc length along a great circle between two points, and returns a numeric value expression in degrees.

For example, a function computing the distance between two points of coordinates (25,-19.5) and (25.4,-20) both expressed according to the ICRS coordinate system with GEOCENTER reference position, would be written as follows:

DISTANCE(POINT('ICRS GEOCENTER',25.0,-19.5),
POINT('ICRS GEOCENTER',25.4, -20.0))

, where all numeric values and the returned arc-length are in degrees.

The distance between two points could also be derived from two columns referencing POINT geometries stored in the database tables as follows:

```
DISTANCE(t.p1,t.p2)
```

, where *t* would be the table and *p1, p2* the column references for the POINT geometries.

Since the two argument points may be expressed in different coordinate systems, the function is responsible for converting one (or both). If it cannot do so, it SHOULD throw an error message, to be defined by the service making use of ADQL.

### 2.4.11 INTERSECTS

This numeric function determines if two geometry values overlap. This is most commonly used to express a "shape-vs-shape" intersection test.

For example, to determine whether a circle of one degree radius centered in a position of (25.4, -20.0) degrees and defined according to the ICRS coordinate system with GEOCENTER reference position overlaps with a box of ten degrees centered in a position (20.0, -15.0) in degrees and defined according to the same coordinate system, we would make use of the INTERSECTS function as follows:

```
INTERSECTS(CIRCLE('ICRS GEOCENTER', 25.4, -20.0, 1),
           BOX('ICRS GEOCENTER', 20.0, -15.0, 10, 10))
```

, where the INTERSECTS function returns 1 (true) if the two arguments overlap and 0 (false) otherwise. When used in the where clause of a query, the value must be compared to 0 or 1 to form an SQL predicate:

```
INTERSECTS(CIRCLE('ICRS GEOCENTER', 25.4, -20.0, 1),
           BOX('ICRS GEOCENTER', 20.0, -15.0, 10, 10)) = 1
```

for "does intersect" and

```
INTERSECTS(CIRCLE('ICRS GEOCENTER', 25.4, -20.0, 1),
           BOX('ICRS GEOCENTER', 20.0, -15.0, 10, 10)) = 0
```

for "does not intersect".

The arguments to the INTERSECTS function can be (literal) values created from the geometry types or they can be single column names or aliases (for geometry stored in a database table). Note that if one of the arguments is a POINT, intersects is equivalent to contains (with the point argument first). Unlike

CONTAINS, the function's arguments are commutative, e.g. INTERSECTS(a, b) is equivalent to INTERSECTS(b, a). Since the two argument points may be expressed in different coordinate systems, the function is responsible for converting one (or both). If it cannot do so, it SHOULD throw an error message, to be defined by the service making use of ADQL.

### 2.4.12 POINT

This function expresses a single location on the sky, and corresponds semantically to an STC SpatialCoord ([3], section 4.4.2). The arguments specify the coordinate system and the position, where:

- the coordinate system is a string value expression as defined in Section 2.4.1.
- the position is a comma separated numeric duple, with units and legal ranges as defined in Section 2.4.1.

For example, a function expressing a point with right ascension of 25 degrees and declination of -19.5 degrees according to the ICRS coordinate system with GEOCENTER reference position, would be written as follows:

POINT('ICRS GEOCENTER', 25.0,-19.5)

, where numeric values are in degrees. The coordinates of the POINT could also be derived from the coordinate's column references:

POINT('ICRS GEOCENTER', t.ra, t.dec)

, where *t* would be the table and *ra*, *dec* the column references for the position.

The coordinates of a POINT could also be individually extracted using the COORD1 and COORD2 functions (See 2.4.7 and 2.4.8).

To see what this function would return when listed in the select clause, see Section 2.4.15.

### 2.4.13 POLYGON

This function expresses a region on the sky with sides denoted by great circles passing through specified coordinates. It corresponds semantically to the STC Polygon region ([3], section 4.5.1.4). The arguments specify the coordinate system and three or more sets of 2-D coordinates, where:

- the coordinate system is a string value expression as defined in Section 2.4.1.
- the coordinate sets are comma separated numeric duples, with units and legal ranges as defined in Section 2.4.1.

For example, a function expressing a triangle, whose vertices are (10.0, -10.5), (20.0, 20.5) and (30.0,30.5) in degrees according to the ICRS coordinate system with GEOCENTER reference position, would be written as follows:

POLYGON('ICRS GEOCENTER', 10.0, -10.5, 20.0, 20.5, 30.0, 30.5)

, where all numeric values are in degrees,

As for other geometries like BOX, CIRCLE and POINT, one could also derive the coordinates from database column references instead:

POLYGON('ICRS GEOCENTER', t.ra, t.dec, 20.0, 20.5, 30.0, 30.5)

, where *t* would be the table and *ra*, *dec* the column references for one of the triangle's corner position.

Thus, the polygon is a list of vertices in a single coordinate system, with each vertex connected to the next along a great circle and the last vertex implicitly connected to the first vertex.

### 2.4.14 REGION

This function provides a generic way of expressing a region represented by a single string input parameter. The format of the string MUST be specified by a service that accepts ADQL by referring to a standard format. Currently STC/s is the only standardized string representation a service can declare.

For example, given a string serialization of an STC region, the REGION function just embeds such literal within parenthesis in the following way:

REGION('Convex … Position … Error … Size')

A detailed description on how to use STC/s can be seen in the referenced document [4]. Inappropriate geometries for this construct SHOULD throw an error message, to be defined by the service making use of ADQL.

### 2.4.15 Geometry in the SELECT clause

Geometry values (literals or columns containing geometry values) may be listed in the select clause, in which case they must be converted into a text form. This text form will be identical to the way a literal value would be specified in a query, including the geometry type (POINT, CIRCLE, BOX, or POLYGON) and all arguments but excluding the required quotes around the coordinate system string. For example, the query

SELECT circle('ICRS GEOCENTER', 1, 2, 0.5)

could return

CIRCLE(‘ICRS GEOCENTER ‘, 1.0, 2.0, 0.5)

or equivalent. The case of the coordinate system string should be preserved; the geometry type string is case insensitive. The output may alter the numeric format by converting whole numbers to floating point (as in the example above) but should not gratuitously add digits. Otherwise, numeric output must conform to the rules for numeric expressions in the ADQL BNF.

## 2.5 User Defined Functions

ADQL also provides a placeholder to define user specific functions. Such construct supports a variable list of parameters as input in the following way:

```
<user_defined_function> ::=
    <user_defined_function_name>
        <left_paren>
            [ <user_defined_function_param> [ { <comma>
                <user_defined_function_param> }... ] ]
        <right_paren>
```

The function names can be qualified with a prefix to ease parsing of the ADQL statement

```
<user_defined_function_name> ::=
    [ <default_function_prefix> ] <regular_identifier>
```

, while the function parameters are generic enough to support string, numeric and geometrical expressions

```
<user_defined_function_param> ::= <value_expression>
```

If metadata on a user defined function is available, this should be used. For example function names and cardinality of arguments should be checked against metadata where available.

# Appendix A: BNF Grammar

An easier to navigate version of the BNF grammar can be found at
http://www.ivoa.net/internal/IVOA/IvoaVOQL/adql-bnf-v2.0.html

```
<ADQL_language_character> ::=
    <simple_Latin_letter>
  | <digit>
  | <SQL_special_character>

<ADQL_reserved_word> ::=
    ABS
  | ACOS
  | AREA
  | ASIN
  | ATAN
  | ATAN2
  | BOX
  | CEILING
  | CENTROID
  | CIRCLE
  | CONTAINS
  | COORD1
  | COORD2
  | COORDSYS
  | COS
  | DEGREES
  | DISTANCE
  | EXP
  | FLOOR
  | INTERSECTS
  | LOG
  | LOG10
  | MOD
  | PI
  | POINT
  | POLYGON
  | POWER
  | RADIANS
  | REGION
  | RAND
  | ROUND
  | SIN
  | SQRT
```

```
  | TOP
  | TAN
  | TRUNCATE

<SQL_embedded_language_character> ::=
    <left_bracket>
  | <right_bracket>

<SQL_reserved_word> ::=
   ABSOLUTE | ACTION | ADD | ALL
 | ALLOCATE | ALTER | AND
 | ANY | ARE
 | AS | ASC
 | ASSERTION | AT
 | AUTHORIZATION | AVG
 | BEGIN | BETWEEN | BIT | BIT_LENGTH
 | BOTH | BY
 | CASCADE | CASCADED | CASE | CAST
 | CATALOG
 | CHAR | CHARACTER | CHAR_LENGTH
 | CHARACTER_LENGTH | CHECK | CLOSE | COALESCE
 | COLLATE | COLLATION
 | COLUMN | COMMIT
 | CONNECT
 | CONNECTION | CONSTRAINT
 | CONSTRAINTS | CONTINUE
 | CONVERT | CORRESPONDING | COUNT | CREATE | CROSS
 | CURRENT
 | CURRENT_DATE | CURRENT_TIME
 | CURRENT_TIMESTAMP | CURRENT_USER | CURSOR
 | DATE | DAY | DEALLOCATE
 | DECIMAL | DECLARE | DEFAULT | DEFERRABLE
 | DEFERRED | DELETE | DESC | DESCRIBE | DESCRIPTOR
 | DIAGNOSTICS
 | DISCONNECT | DISTINCT | DOMAIN | DOUBLE | DROP
 | ELSE | END | END-EXEC | ESCAPE
 | EXCEPT | EXCEPTION
 | EXEC | EXECUTE | EXISTS
 | EXTERNAL | EXTRACT
 | FALSE | FETCH | FIRST | FLOAT | FOR
 | FOREIGN | FOUND | FROM | FULL
 | GET | GLOBAL | GO | GOTO
 | GRANT | GROUP
 | HAVING | HOUR
 | IDENTITY | IMMEDIATE | IN | INDICATOR
 | INITIALLY | INNER | INPUT
```

| INSENSITIVE | INSERT | INT | INTEGER | INTERSECT
| INTERVAL | INTO | IS
| ISOLATION
| JOIN
| KEY
| LANGUAGE | LAST | LEADING | LEFT
| LEVEL | LIKE | LOCAL | LOWER
| MATCH | MAX | MIN | MINUTE | MODULE
| MONTH
| NAMES | NATIONAL | NATURAL | NCHAR | NEXT | NO
| NOT | NULL
| NULLIF | NUMERIC
| OCTET_LENGTH | OF
| ON | ONLY | OPEN | OPTION | OR
| ORDER | OUTER
| OUTPUT | OVERLAPS
| PAD | PARTIAL | POSITION | PRECISION | PREPARE
| PRESERVE | PRIMARY
| PRIOR | PRIVILEGES | PROCEDURE | PUBLIC
| READ | REAL | REFERENCES | RELATIVE | RESTRICT
| REVOKE | RIGHT
| ROLLBACK | ROWS
| SCHEMA | SCROLL | SECOND | SECTION
| SELECT
| SESSION | SESSION_USER | SET
| SIZE | SMALLINT | SOME | SPACE | SQL | SQLCODE
| SQLERROR | SQLSTATE
| SUBSTRING | SUM | SYSTEM_USER
| TABLE | TEMPORARY
| THEN | TIME | TIMESTAMP
| TIMEZONE_HOUR | TIMEZONE_MINUTE
| TO | TRAILING | TRANSACTION
| TRANSLATE | TRANSLATION | TRIM | TRUE
| UNION | UNIQUE | UNKNOWN | UPDATE | UPPER | USAGE
| USER | USING
| VALUE | VALUES | VARCHAR | VARYING | VIEW
| WHEN | WHENEVER | WHERE | WITH | WORK | WRITE
| YEAR
| ZONE

<SQL_special_character> ::=
<space>
| <double_quote>
| <percent>
| <ampersand>
| <quote>

```
 | <left_paren>
 | <right_paren>
 | <asterisk>
 | <plus_sign>
 | <comma>
 | <minus_sign>
 | <period>
 | <solidus>
 | <colon>
 | <semicolon>
 | <less_than_operator>
 | <equals_operator>
 | <greater_than_operator>
 | <question_mark>
 | <underscore>
 | <vertical_bar>

<ampersand> ::= &

<approximate_numeric_literal> ::= <mantissa>E<exponent>

<area> ::= AREA <left_paren> <geometry_value_expression> <right_paren>

<as_clause> ::= [ AS ] <column_name>

<asterisk> ::= *

<between_predicate> ::=
   <value_expression> [ NOT ] BETWEEN
   <value_expression> AND <value_expression>

<boolean_factor> ::= [ NOT ] <boolean_primary>

<boolean_primary> ::=
   | <left_paren> <search_condition> <right_paren>
     <predicate>

<boolean_term> ::=
    <boolean_factor>
  | <boolean_term> AND <boolean_factor>

<box> ::=
  BOX <left_paren> <coord_sys>
           <comma> <coordinates>
           <comma> <numeric_value_expression>
           <comma> <numeric_value_expression>
```

<right_paren>

<catalog_name> ::= <identifier>

<centroid> ::= CENTROID <left_paren> <geometry_value_expression>
<right_paren>

<character_factor> ::= <character_primary>

<character_primary> ::=
<value_expression_primary>
| <string_value_function>

<character_representation> ::= <nonquote_character> | <quote_symbol>

<character_string_literal> ::=
<quote> [ <character_representation>... ] <quote>
[ { <separator>... <quote> [ <character_representation>... ] <quote> }... ]

<character_value_expression> ::= <concatenation> | <character_factor>

<circle> ::=
CIRCLE <left_paren> <coord_sys>
<comma> <coordinates>
<comma> <radius>
<right_paren>

<colon> ::= :

<column_name> ::= <identifier>

<column_name_list> ::= <column_name> [ { <comma> <column_name> }... ]

<column_reference> ::= [ <qualifier> <period> ] <column_name>

<comma> ::= ,

<comment> ::= <comment_introducer> [ <comment_character>... ] <newline>

<comment_character> ::= <nonquote_character> | <quote>

<comment_introducer> ::= <minus_sign><minus_sign> [<minus_sign>...]

<comp_op> ::=
<equals_operator>
| <not_equals_operator>

| <less_than_operator>
| <greater_than_operator>
| <less_than_or_equals_operator>
| <greater_than_or_equals_operator>

<comparison_predicate> ::=
<value_expression> <comp_op> <value_expression>

<concatenation> ::= <character_value_expression> <concatenation_operator>
<character_factor>

<concatenation_operator> ::= ||

<contains> ::= CONTAINS <left_paren>
<geometry_value_expression> <comma> <geometry_value_expression>
<right_paren>

<coord1> ::= COORD1 <left_paren> <coord_value> <right_paren>

<coord2> ::= COORD2 <left_paren> <coord_value> <right_paren>

<coord_sys> ::= <string_value_expression>

<coord_value> ::= <point> | <column_reference>

<coordinate1> ::= <numeric_value_expression>

<coordinate2> ::= <numeric_value_expression>

<coordinates> ::= <coordinate1> <comma> <coordinate2>

<correlation_name> ::= <identifier>

<correlation_specification> ::= [ AS ] <correlation_name>

<default_function_prefix> ::=

<delimited_identifier> ::= <double_quote> <delimited_identifier_body>
<double_quote>

<delimited_identifier_body> ::= <delimited_identifier_part>...

<delimited_identifier_part> ::= <nondoublequote_character> |
<double_quote_symbol>

<delimiter_token> ::=

```
    <character_string_literal>
  | <delimited_identifier>
  | <SQL_special_character>
  | <not_equals_operator>
  | <greater_than_or_equals_operator>
  | <less_than_or_equals_operator>
  | <concatenation_operator>
  | <double_period>
  | <left_bracket>
  | <right_bracket>

<derived_column> ::= <value_expression> [ <as_clause> ]

<derived_table> ::= <table_subquery>

<digit> ::= 0 | 1 | 2 | 3 | 4 | 5 | 6 | 7 | 8 | 9

<distance_function> ::=
  DISTANCE <left_paren> <coord_value> <comma> <coord_value>
    <right_paren>

<double_period> ::= ..

<double_quote> ::= "

<double_quote_symbol> ::= <double_quote><double_quote>

<equals_operator> ::= =

<exact_numeric_literal> ::=
   <unsigned_integer> [ <period> [ <unsigned_integer> ] ]
  | <period> <unsigned_integer>

<exists_predicate> ::= EXISTS <table_subquery>

<exponent> ::= <signed_integer>

<extract_coordsys> ::= COORDSYS <left_paren>
   <geometry_value_expression>
  <right_paren>

<factor> ::= [ <sign> ] <numeric_primary>

<from_clause> ::= FROM <table_reference>
  [ { <comma> <table_reference> }... ]
```

<general_literal> ::= <character_string_literal>

<general_set_function> ::=
<set_function_type> <left_paren> [ <set_quantifier> ] <value_expression>
<right_paren>

<geometry_value_expression> ::= <value_expression_primary > |
<geometry_value_function>

<geometry_value_function> ::=
<box>
| <centroid>
| <circle>
| <point>
| <polygon>
| <region>

<greater_than_operator> ::= >

<greater_than_or_equals_operator> ::= >=

<group_by_clause> ::= GROUP BY <grouping_column_reference_list>

<grouping_column_reference> ::= <column_reference>

<grouping_column_reference_list> ::=
<grouping_column_reference> [ { <comma> <grouping_column_reference>
}... ]

<having_clause> ::= HAVING <search_condition>

<identifier> ::= <regular_identifier> | <delimited_identifier>

<in_predicate> ::=
<value_expression> [ NOT ] IN <in_predicate_value>

<in_predicate_value> ::=
<table_subquery> | <left_paren> <in_value_list> <right_paren>

<in_value_list> ::=
<value_expression> { <comma> <value_expression> } ...

<intersects > ::= INTERSECTS <left_paren> <geometry_value_expression>
<comma> <geometry_value_expression> <right_paren>

<join_column_list> ::= <column_name_list>

```
<join_condition> ::= ON <search_condition>

<join_specification> ::= <join_condition> | <named_columns_join>

<join_type> ::=
  INNER
 | <outer_join_type> [ OUTER ]

<joined_table> ::=
   <qualified_join>
  | <left_paren> <joined_table> <right_paren>

<keyword> ::= <SQL_reserved_word> | <ADQL_reserved_word>

<left_bracket> ::= [

<left_paren> ::= (

<less_than_operator> ::= <

<less_than_or_equals_operator> ::= <=

<like_predicate> ::=
   <match_value> [ NOT ] LIKE <pattern>

<mantissa> ::= <exact_numeric_literal>

<match_value> ::= <character_value_expression>

<math_function> ::=
   ABS <left_paren> <numeric_value_expression> <right_paren>
        | CEILING <left_paren> <numeric_value_expression> <right_paren>
        | DEGREES <left_paren> <numeric_value_expression> <right_paren>
        | EXP <left_paren> <numeric_value_expression> <right_paren>
        | FLOOR <left_paren> <numeric_value_expression> <right_paren>
        | LOG <left_paren> <numeric_value_expression> <right_paren>
        | LOG10 <left_paren> <numeric_value_expression> <right_paren>
        | MOD <left_paren> <numeric_value_expression> <comma>
           <numeric_value_expression> <right_paren>
        | PI <left_paren><right_paren>
        | POWER <left_paren> <numeric_value_expression> <comma>
           <numeric_value_expression> > <right_paren>
        | RADIANS <left_paren> <numeric_value_expression> <right_paren>
        | RAND <left_paren> [ <unsigned_integer> ] <right_paren>
        | ROUND <left_paren> <numeric_value_expression>
```

[ <comma> <signed_integer>] <right_paren>
| SQRT <left_paren> <numeric_value_expression> <right_paren>
| TRUNCATE <left_paren> <numeric_value_expression>
[ <comma> <signed_integer>] <right_paren>

<minus_sign> ::= -

<named_columns_join> ::= USING <left_paren> <join_column_list>
<right_paren>

<newline> ::=

<non_predicate_geometry_function> ::=
<area>
| <coord1>
| <coord2>
| <distance>

<nondelimiter_token> ::=
<regular_identifier>
| <keyword>
| <unsigned_numeric_literal>

<nondoublequote_character> ::=

<nonquote_character> ::=

<not_equals_operator> ::= <not_equals_operator1> | <not_equals_operator2>

<not_equals_operator1> ::= <>

<not_equals_operator2> ::= !=

<null_predicate> ::= <column_reference> IS [ NOT ] NULL

<numeric_geometry_function> ::=
<predicate_geometry_function> | <non_predicate_geometry_function>

<numeric_primary> ::=
<value_expression_primary>
| <numeric_value_function>

<numeric_value_expression> ::=
<term>
| <numeric_value_expression> <plus_sign> <term>
| <numeric_value_expression> <minus_sign> <term>

<numeric_value_function> ::=
<trig_function>
| <math_function>
| <numeric_geometry_function >
| <user_defined_function>

<order_by_clause> ::= ORDER BY <sort_specification_list>

<ordering_specification> ::= ASC | DESC

<outer_join_type> ::= LEFT | RIGHT | FULL

<pattern> ::= <character_value_expression>

<percent> ::= %

<period> ::= .

<plus_sign> ::= +

<point> ::= POINT <left_paren> <coord_sys> <comma> <coordinates>
<right_paren>

<polygon> ::=
POLYGON <left_paren> <coord_sys>
<comma> <coordinates>
<comma> <coordinates>
{ <comma> <coordinates> } ?
<right_paren>

<predicate> ::=
<comparison_predicate>
| <between_predicate>
| <in_predicate>
| <like_predicate>
| <null_predicate>
| <exists_predicate>

<predicate_geometry_function> ::= <contains> | <intersects>

<qualified_join> ::=
<table_reference> [ NATURAL ] [ <join_type> ] JOIN
<table_reference> [ <join_specification> ]

<qualifier> ::= <table_name> | <correlation_name>

```
 <query_expression> ::=
    <query_specification>
  | <joined_table>

<query_specification> ::=
   SELECT [ <set_quantifier> ] [ <set_limit> ] <select_list> <table_expression>

<question_mark> ::= ?

<quote> ::= '

<quote_symbol> ::= <quote> <quote>

<radius> ::= <numeric_value_expression>

<region> ::= REGION <left_paren> <string_value_expression> <right_paren>

<regular_identifier> ::=
   <simple_Latin_letter>...
   [ { <digit> | <simple_Latin_letter> | <underscore> }... ]

<right_bracket> ::= ]

<right_paren> ::= )

<schema_name> ::= [ <catalog_name> <period> ] <unqualified_schema name>

<search_condition> ::=
   <boolean_term>
 | <search_condition> OR <boolean_term>

<select_list> ::=
   <asterisk>
 | <select_sublist> [ { <comma> <select_sublist> }... ]

<select_sublist> ::= <derived_column> | <qualifier> <period> <asterisk>

<semicolon> ::= ;

<separator> ::= { <comment> | <space> | <newline> }...

<set_function_specification> ::=
   COUNT <left_paren> <asterisk> <right_paren>
 | <general_set_function>
```

<set_function_type> ::= AVG | MAX | MIN | SUM | COUNT

<set_limit> ::= TOP <unsigned_integer>

<set_quantifier> ::= DISTINCT | ALL

<sign> ::= <plus_sign> | <minus_sign>

<signed_integer> ::= [ <sign> ] <unsigned_integer>

<simple_Latin_letter> ::=
<simple_Latin_upper_case_letter>
| <simple_Latin_lower_case_letter>

<simple_Latin_lower_case_letter> ::=
a | b | c | d | e | f | g | h | i | j | k | l | m | n | o
| p | q | r | s | t | u | v | w | x | y | z

<simple_Latin_upper_case_letter> ::=
A | B | C | D | E | F | G | H | I | J | K | L | M | N | O
| P | Q | R | S | T | U | V | W | X | Y | Z

<solidus> ::= /

<sort_key> ::= <column_name> | <unsigned_integer>

<sort_specification> ::=
<sort_key> [ <ordering_specification> ]

<sort_specification_list> ::=
<sort_specification> [ { <comma> <sort_specification> }... ]

<space> ::=

<string_geometry_function> ::= <extract_coordsys>

<string_value_expression> ::= <character_value_expression>

<string_value_function> ::= <string_geometry_function> |
<user_defined_function>

<subquery> ::= <left_paren> <query_expression> <right_paren>

<table_expression> ::=
<from_clause>
[ <where_clause> ]

```
    [ <group_by_clause> ]
    [ <having_clause> ]
    [ <order_by_clause> ]

<table_name> ::= [ <schema_name> <period> ] <identifier>

<table_reference> ::=
   <table_name> [ <correlation_specification> ]
 | <derived_table> <correlation_specification>
 | <joined_table>

<table_subquery> ::= <subquery>

<term> ::=
    <factor>
  | <term> <asterisk> <factor>
  | <term> <solidus> <factor>

<token> ::=
    <nondelimiter_token>
  | <delimiter_token>

 <trig_function> ::=
     ACOS <left_paren> <numeric_value_expression> <right_paren>
   | ASIN <left_paren> <numeric_value_expression> <right_paren>
   | ATAN <left_paren> <numeric_value_expression> <right_paren>
   | ATAN2 <left_paren> <numeric_value_expression> <comma>
       <numeric_value_expression> <right_paren>
   | COS <left_paren> <numeric_value_expression> <right_paren>
   | COT <left_paren> <numeric_value_expression> <right_paren>
   | SIN <left_paren> <numeric_value_expression> <right_paren>
   | TAN <left_paren> <numeric_value_expression> <right_paren>

<underscore> ::= _

<unqualified_schema name> ::= <identifier>

<unsigned_integer> ::= <digit>...

<unsigned_literal> ::= <unsigned_numeric_literal> | <general_literal>

<unsigned_numeric_literal> ::=
   <exact_numeric_literal>
 | <approximate_numeric_literal>

<unsigned_value_specification> ::= <unsigned_literal>
```

```
<user_defined_function> ::=
    <user_defined_function_name>
        <left_paren>
            [ <user_defined_function_param> [ { <comma>
              <user_defined_function_param> }... ] ]
        <right_paren>

<user_defined_function_name> ::=
     [ <default_function_prefix> ] <regular_identifier>

<user_defined_function_param> ::= <value_expression>

<value_expression> ::=
    <numeric_value_expression>
  | <string_value_expression>
  | <geometry_value_expression>

<value_expression_primary> ::=
    <unsigned_value_specification>
  | <column_reference>
  | <set_function_specification>
  | <left_paren> <value_expression> <right_paren>

<vertical_bar> ::= |

<where_clause> ::= WHERE <search_condition>
```